\documentclass[%
reprint,
showpacs,preprintnumbers,
nofootinbib,
amsmath,amssymb,
aps,
prd,
floatfix,
]{revtex4-1}
\usepackage{graphicx}
\usepackage{dcolumn}
\usepackage{bm}
\usepackage{url}
\usepackage[ddmmyy,24hr]{datetime}
\usepackage{color}

\newcommand{\nua}[1]{\ensuremath{\rlap
           {\kern-2.5pt\ensuremath
           {\overset{\scriptscriptstyle(-)}{\phantom{\nu}}}}
           {\ensuremath{{\nu}_{#1}}}}}
\begin{document}

\preprint{\begin{tabular}{l}
\texttt{arXiv:1212.3805 [hep-ph]}
\end{tabular}}

\title{Short-Baseline Electron Neutrino Oscillation Length After Troitsk}

\author{C. Giunti}
\affiliation{INFN, Sezione di Torino, Via P. Giuria 1, I--10125 Torino, Italy}

\author{M. Laveder}
\affiliation{Dipartimento di Fisica e Astronomia ``G. Galilei'', Universit\`a di Padova,
and
INFN, Sezione di Padova,
Via F. Marzolo 8, I--35131 Padova, Italy}

\author{Y.F. Li}
\affiliation{Institute of High Energy Physics,
Chinese Academy of Sciences, Beijing 100049, China}

\author{H.W. Long}
\affiliation{Department of Modern Physics, University of Science and
Technology of China, Hefei, Anhui 230026, China}


\begin{abstract}
We discuss the implications
for short-baseline electron neutrino disappearance in the 3+1 mixing scheme
of the recent Troitsk
bounds on the mixing of a neutrino with mass between 2 and 100 eV.
Considering the Troitsk data in combination with the
results of short-baseline $\nu_{e}$ and $\bar\nu_{e}$ disappearance experiments,
which include the reactor and Gallium anomalies,
we derive a $2\sigma$ allowed range for the effective neutrino squared-mass difference
between $0.85$ and $ 43 \, \text{eV}^2$.
The upper bound implies that it is likely that oscillations
in distance and/or energy can be observed
in radioactive source experiments.
It is also favorable for the ICARUS@CERN experiment,
in which it is likely that oscillations are not washed-out in the near detector.
We discuss also the implications for neutrinoless double-$\beta$ decay.
\end{abstract}


\maketitle

The reactor $\bar\nu_{e}$
\cite{Mueller:2011nm,Huber:2011wv,Mention:2011rk}
and
the Gallium $\nu_{e}$
\cite{Laveder:2007zz,hep-ph/0610352,Giunti:2010zu}
anomalies indicate that electron neutrino and antineutrinos
may disappear at short distances
because of oscillations generated by a squared-mass difference $\Delta{m}^2_{\text{SBL}} \gtrsim 1 \, \text{eV}^2$
\cite{Laveder:2007zz,hep-ph/0610352,Giunti:2010zu,1210.5715}.
Since $\Delta{m}^2_{\text{SBL}}$ is much larger than the two $\Delta{m}^2$'s
which generate the observed solar, atmospheric and long-baseline oscillations
in standard three-neutrino mixing
(see \cite{Giunti:2007ry,Bilenky:2010zza,Xing:2011zza}),
we are led to consider the so-called 3+1 neutrino mixing scheme,
which is the extension of standard three-neutrino mixing with an additional massive neutrino.
This is the simplest extension of standard three-neutrino mixing
which can explain the reactor and Gallium anomalies.
In the flavor basis, the additional neutrino is sterile,
because from the LEP measurement of the
invisible width of the $Z$ boson
\cite{hep-ex/0509008}
we know that there are only three light active flavor neutrinos,
$\nu_{e}$,
$\nu_{\mu}$ and
$\nu_{\tau}$.
Hence,
we have the mixing relation
\begin{equation}
\nu_{\alpha}
=
\sum_{k=1}^{4} U_{\alpha k} \nu_{k}
\qquad
(\alpha=e,\mu,\tau,s)
\,,
\label{mixing}
\end{equation}
between the flavor fields $\nu_{\alpha}$ ($\nu_{s}$ is the sterile neutrino) and the massive fields $\nu_{k}$,
with respective masses $m_{k}$.
$U$ is the unitary $4\times4$ mixing matrix.
The effective survival probability
at a distance $L$ of electron neutrinos and antineutrinos with energy $E$
in short-baseline (SBL) neutrino oscillation experiments
is given by
(see Refs.~\cite{hep-ph/9812360,hep-ph/0405172,hep-ph/0606054,GonzalezGarcia:2007ib})
\begin{equation}
P_{\nua{e}\to\nua{e}}^{\text{SBL}}
=
1
-
\sin^{2}2\vartheta_{ee}
\sin^{2}\left( \frac{\Delta{m}^2_{41} L}{4E} \right)
\,,
\label{survi}
\end{equation}
with
$\Delta{m}^2_{41} \equiv m_{4}^2 - m_{1}^2 = \Delta{m}^2_{\text{SBL}}$
and
the transition amplitude
\begin{equation}
\sin^{2}2\vartheta_{ee}
=
4 |U_{e4}|^2 \left( 1 - |U_{e4}|^2 \right)
\,.
\label{survisin}
\end{equation}

\begin{figure}[t]
\begin{center}
\includegraphics*[width=\linewidth]{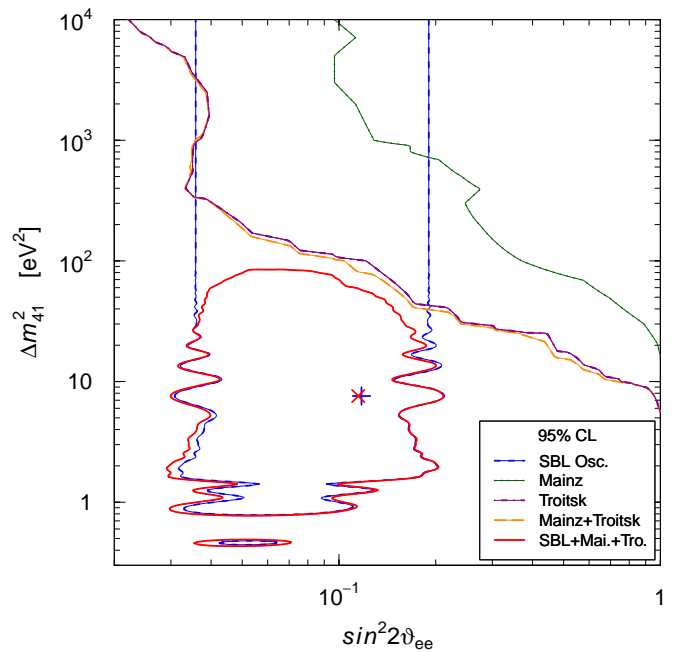}
\end{center}
\caption{ \label{fig:sup-osc-mnz-tsk}
Comparison of the 95\% CL
allowed region in the
$\sin^{2}2\vartheta_{ee}$--$\Delta{m}^{2}_{41}$ plane
obtained from the global fit of $\nu_{e}$ and $\bar\nu_{e}$ short-baseline oscillation data
\cite{1210.5715},
the 95\% CL bounds obtained from
Mainz
\cite{1210.4194}
and
Troitsk
\cite{1211.7193}
data,
and
the allowed region obtained from the combined fit.
The best-fit points of the oscillation and combined analyses are indicated, respectively, by ``$+$'' and ``$\times$''.
}
\end{figure}

In Ref.~\cite{1210.5715} we presented an update of
the 3+1 analysis of short-baseline $\nu_{e}$ and $\bar\nu_{e}$ disappearance experiments\footnote{
We do not consider here, as well as in Ref.~\cite{1210.5715},
the more controversial
LSND
\cite{hep-ex/0104049}
and
MiniBooNE
\cite{1207.4809}
$\nua{\mu}\to\nua{e}$
anomalies,
whose explanation in the framework of neutrino oscillations
is problematic
(see Refs.~\cite{1103.4570,1107.1452,1109.4033,1111.1069,1207.4765}).}
which took into account:
1)
the data of the
Bugey-3 \cite{Declais:1995su},
Bugey-4 \cite{Declais:1994ma},
ROVNO91 \cite{Kuvshinnikov:1990ry},
Gosgen \cite{Zacek:1986cu},
ILL \cite{Hoummada:1995zz}
and
Krasnoyarsk \cite{Vidyakin:1990iz}
reactor antineutrino experiments,
with the new theoretical fluxes
\cite{Mueller:2011nm,Huber:2011wv,Mention:2011rk};
2)
the data of the
GALLEX
\cite{Anselmann:1995ar,Hampel:1998fc,1001.2731}
and
SAGE
\cite{Abdurashitov:1996dp,hep-ph/9803418,nucl-ex/0512041,0901.2200}
Gallium radioactive source experiments
with the statistical method discussed in Ref.~\cite{Giunti:2010zu},
considering the recent
${}^{71}\text{Ga}({}^{3}\text{He},{}^{3}\text{H}){}^{71}\text{Ge}$
cross section measurement in Ref.~\cite{Frekers:2011zz};
3)
the solar neutrino constraint on $\sin^{2}2\vartheta_{ee}$
\cite{Giunti:2009xz,Palazzo:2011rj,Palazzo:2012yf,Palazzo-NOW2012,1210.5715};
4)
the
KARMEN \cite{Bodmann:1994py,hep-ex/9801007}
and
LSND \cite{hep-ex/0105068}
$\nu_{e} + {}^{12}\text{C} \to {}^{12}\text{N}_{\text{g.s.}} + e^{-}$
scattering data \cite{1106.5552},
with the method discussed in Ref.~\cite{Giunti:2011cp}.
In Fig.~\ref{fig:sup-osc-mnz-tsk}
we reproduce the allowed 95\% CL region
in the $\sin^{2}2\vartheta_{ee}$--$\Delta{m}^2_{41}$
plane presented in Ref.~\cite{1210.5715}.
One can see that
there is no upper limit for
$\Delta{m}^2_{41}$
from oscillation data.
In Ref.~\cite{1210.5715}
we discussed the possibilities to constrain
$\sin^{2}2\vartheta_{ee}$
and
$\Delta{m}^2_{41}$
with
measurements of the effects of $m_4$
on the electron spectrum in $\beta$-decay far from the end-point
and
with neutrinoless double-$\beta$ decay
(if massive neutrinos are Majorana particles)\footnote{
Let us only mention that
cosmological measurements give information on
the number of neutrinos
and on the values of neutrino masses at the eV scale
(see \cite{Wong:2011ip,Abazajian:2012ys,Steigman:2012ve}),
but the results depend on the theoretical assumption of a cosmological model.
}
assuming the natural mass hierarchy
\begin{equation}
m_{1},m_{2},m_{3} \ll m_{4}
\,,
\label{mass-hierarchy}
\end{equation}
which implies
\begin{equation}
m_{4}^2 \simeq \Delta{m}^2_{41}
\,.
\label{m4}
\end{equation}
In particular,
we showed that the recent
Tritium $\beta$-decay
data of the Mainz Neutrino Mass Experiment
\cite{1210.4194}
constrain $\Delta{m}^2_{41}$ to be smaller than about $10^{4}\,\text{eV}^2$
at 95\% CL.

Recently the Troitsk collaboration presented the results
\cite{1211.7193}
of a search for the effects of $m_{4}^2$
between 4 and $10^{4}\,\text{eV}^2$
in the spectrum of the electrons emitted in Tritium decay
in the Troitsk nu-mass experiment.
Since they did not find any deviation from the massless neutrino case,
their data allowed them to constrain the value of
$|U_{e4}|^2$
as a function of $m_{4}^2$
in a similar way as done by the Mainz collaboration.
Since the Troitsk bounds are significantly stronger than the Mainz bounds,
in this paper we present an update of the analysis in Ref.~\cite{1210.5715}
which takes into account the Troitsk data.

Figure~\ref{fig:sup-osc-mnz-tsk}
shows the 95\% CL exclusion curves in the
$\sin^{2}2\vartheta_{ee}$--$\Delta{m}^2_{41}$
plane obtained with the Mainz and Troitsk data.
One can see that the Troitsk exclusion curve cuts the region allowed at 95\% CL
by short-baseline oscillation data
for
values of $\Delta{m}^2_{41}$
between about $40 \, \text{eV}^2$
and $400 \, \text{eV}^2$.
In this interval of $\Delta{m}^2_{41}$
the Troitsk upper bound on $\sin^{2}2\vartheta_{ee}$
is from about four to six times more stringent 
than that of Mainz.
For completeness, in Fig.~\ref{fig:sup-osc-mnz-tsk}
we have shown also the combined Mainz and Troitsk exclusion curve,
but one can see that the improvement with respect to the Troitsk exclusion curve
is very small.

\begin{figure}[t!]
\begin{center}
\includegraphics*[width=\linewidth]{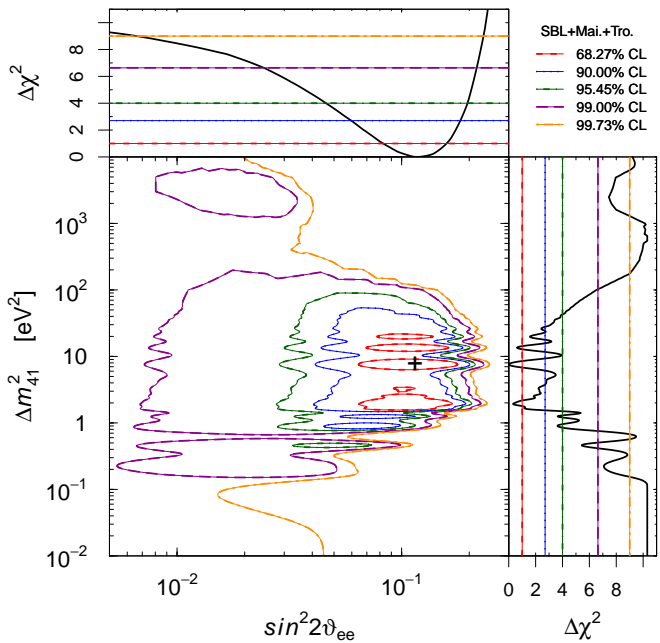}
\end{center}
\caption{ \label{fig:cnt-osc-mnz-tsk}
Allowed regions in the
$\sin^{2}2\vartheta_{ee}$--$\Delta{m}^{2}_{41}$ plane
and
marginal $\Delta\chi^{2}$'s
for
$\sin^{2}2\vartheta_{ee}$ and $\Delta{m}^{2}_{41}$
obtained from the combined fit of $\nu_{e}$ and $\bar\nu_{e}$ short-baseline oscillation data
and the data of the
Mainz
\cite{1210.4194}
and
Troitsk
\cite{1211.7193}
experiments.
The best-fit point is indicated by a ``$+$''.
}
\end{figure}

\begin{figure}[t]
\begin{center}
\includegraphics*[width=\linewidth]{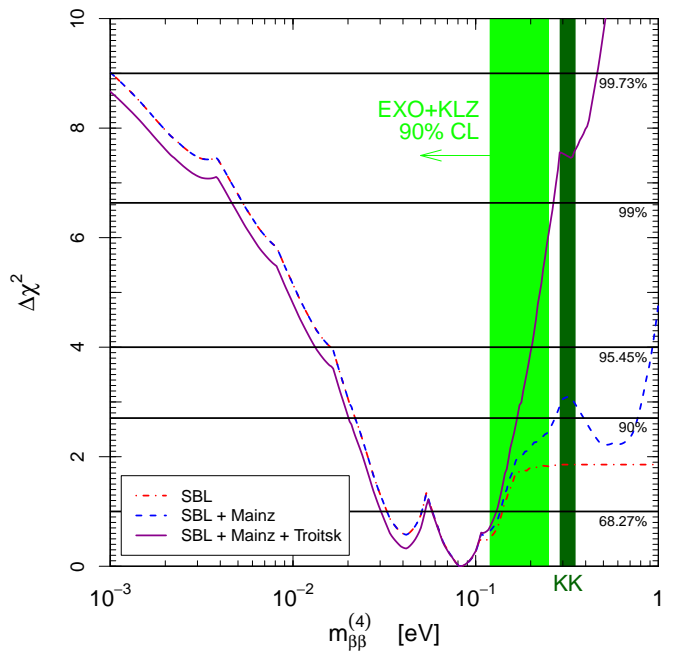}
\end{center}
\caption{ \label{fig:mbb-chi}
Marginal
$\Delta\chi^2$
as a function of
$m_{\beta\beta}^{(4)}$
obtained from the fit of $\nu_{e}$ and $\bar\nu_{e}$ short-baseline oscillation data (dash-dotted curve),
from the combined fit of oscillation and Mainz data (dashed curve),
and
from the combined fit of oscillation, Mainz and Troitsk data (solid curve).
The vertical green band represents the currently most stringent upper bound for $m_{\beta\beta}^{(4)}$
in the no-cancellation case (see text)
given by the combined
EXO
and
KamLAND-Zen
90\% CL bound on $m_{\beta\beta}$ taking into account nuclear matrix element uncertainties \cite{1211.3863}.
The vertical dark-green band corresponds to the $1\sigma$
Klapdor-Kleingrothaus et al. range of $m_{\beta\beta}$
\cite{KlapdorKleingrothaus:2006ff}.
}
\end{figure}

The allowed region in Fig.~\ref{fig:sup-osc-mnz-tsk}
obtained from the combined fit of
short-baseline oscillation data
and
Mainz
and
Troitsk
data
shows that the value of $\Delta{m}^2_{41}$ is bounded from above.
Figure~\ref{fig:cnt-osc-mnz-tsk}
shows the combined allowed regions
in the
$\sin^{2}2\vartheta_{ee}$--$\Delta{m}^2_{41}$
plane at different CL's
and the marginal $\Delta\chi^{2}=\chi^2-\chi^2_{\text{min}}$'s
for
$\sin^{2}2\vartheta_{ee}$ and $\Delta{m}^{2}_{41}$.

In order to get an estimate of the allowed range of
$\Delta{m}^{2}_{41}$
we consider the corresponding marginal $\Delta\chi^{2}$,
which gives
\begin{equation}
0.85
\lesssim
\Delta{m}^{2}_{41}
\lesssim
 43
\, \text{eV}^2
\qquad
(2\sigma)
\,.
\label{dm2bnd}
\end{equation}
This is a very interested range, because it implies that
the oscillation length
$
L^{\text{osc}}_{41}
=
4 \pi E / \Delta{m}^{2}_{41}
$
is in the interval
\begin{equation}
6 \, \text{cm}
\lesssim
\frac{L^{\text{osc}}_{41}}{E\,[\text{MeV}]}
\lesssim
3
\, \text{m}
\qquad
(2\sigma)
\,.
\label{Losc}
\end{equation}

Taking into account that
electron neutrino and antineutrino radioactive sources
have a typical size of a few centimeters,
there are good possibilities that new experiments
with these sources
\cite{hep-ex/9901012,1006.2103,1011.4509,1107.2335,1109.6036,1110.2983,Abazajian:2012ys,Ianni-NOW2012,Link-NOW2012}
can measure the dependence of the disappearance probability
as a function of distance and/or energy.
Such a measurement will be a smoking-gun proof of
short-baseline oscillations and of the existence of light sterile neutrinos.

For radioactive source experiments using electron neutrinos produced by electron capture,
which have a discrete spectrum,
the oscillatory pattern of the survival probability can be observed if the detector has
a spatial resolution which is much smaller than the oscillation length,
i.e. much smaller than
$6 \, \text{cm}$
if $L^{\text{osc}}_{41}$ is close to the lower bound in Eq.~(\ref{Losc}).
In this case,
the sensitivity to oscillations could be enhanced by using a very thin source
in one or two spatial dimensions
(for example a long and thin cylinder or a flat rectangular parallelepiped).
Then,
if $L^{\text{osc}}_{41}$ is close to the lower bound in Eq.~(\ref{Losc})
one could observe an interesting three-dimensional pattern in
which the averaged survival probability oscillates as a function of distance
along the thin direction(s)
and
does not depend on distance along the thick direction(s).

Experiments using radioactive $\beta^{-}$ sources of electron antineutrinos
with a continuous spectrum
can measure also the oscillatory pattern of the survival probability
as a function of energy
if the spatial resolution of the detector is smaller than the oscillation length
and if the energy resolution
$\Delta E$
is such that
\begin{equation}
\Delta{E} / E
\ll
L^{\text{osc}}_{41} / L
\,.
\label{energyresolution}
\end{equation}
Since electron antineutrinos are detected with the inverse neutron decay reaction
$\bar\nu_{e} + p \to n + e^{+}$
with a threshold of 1.8 MeV,
if $L^{\text{osc}}_{41}$ is close to the lower bound in Eq.~(\ref{Losc})
a spatial resolution much smaller than 10 cm
and
an energy resolution much better than 10\% at $L \sim 1 \, \text{m}$
are needed.

The proposed ICARUS@CERN experiment
\cite{Antonello:2012hf,Stanco-NOW2012,Gibin-NOW2012}
is based on two Liquid-Argon Time-Projection-Chamber imaging detectors
at 300 m and 1.6 km from the source.
Since the beam will have an average neutrino energy of about 2 GeV,
the oscillation length is larger than the distance of the near detector for
$\Delta{m}^{2}_{41} \lesssim 20 \, \text{eV}^2$.
This upper bound is of the same order than that obtained in Eq.~(\ref{dm2bnd}).
Hence, our results imply that the ICARUS@CERN experiment
has good possibilities to measure the disappearance of electron neutrinos
if the oscillation interpretation of the reactor and Gallium anomalies is correct,
because oscillations are not washed-out in the near detector.

There are also several projects aimed at the measurement of the short-baseline disappearance
reactor $\bar\nu_{e}$'s
\cite{1204.2449,1205.2955,Abazajian:2012ys,Gaffiot-NOW2012,1212.2182}.
Taking into account that the product of the reactor $\bar\nu_{e}$ flux and the detection cross section
peaks at about 4 MeV,
the size of a research reactor is about 50 cm,
and a detector cannot be placed closer than a few meters from a reactor,
the distance and/or energy dependence of the survival probability may be measured if the upper bound on
$\Delta{m}^{2}_{41}$
is about an order of magnitude smaller than that in Eq.~(\ref{dm2bnd}).
Such restriction\footnote{
Similar considerations apply to the IsoDAR proposal in Ref.~\cite{1205.4419}.
}
may come from the results of the KATRIN experiment
\cite{Riis:2010zm,SejersenRiis:2011sj,Formaggio:2011jg,Esmaili:2012vg}.

Let us also notice that the marginal $\Delta\chi^{2}$ for $\sin^{2}2\vartheta_{ee}$
in Fig.~\ref{fig:cnt-osc-mnz-tsk}
is similar to that obtained in Ref.~\cite{1210.5715}
from short-baseline data alone.
It gives the interesting interval
\begin{equation}
0.05
\lesssim
\sin^{2}2\vartheta_{ee}
\lesssim
0.19
\qquad
(2\sigma)
\,,
\label{st2bnd}
\end{equation}
which is testable in future short-baseline experiments
with
electron neutrino and antineutrino radioactive sources
\cite{hep-ex/9901012,1006.2103,1011.4509,1107.2335,1109.6036,1110.2983,Abazajian:2012ys,Ianni-NOW2012,Link-NOW2012,1205.4419},
reactor electron antineutrinos
\cite{1204.2449,1205.2955,Abazajian:2012ys,Gaffiot-NOW2012,1212.2182}
and
accelerator electron neutrinos
\cite{Antonello:2012hf,Stanco-NOW2012,Gibin-NOW2012}.

The bounds on $\sin^{2}2\vartheta_{ee}$ and $\Delta{m}^{2}_{41}$ that we have obtained
allow us to update the predictions of Ref.~\cite{1210.5715} for the
contribution
$m_{\beta\beta}^{(4)} \simeq |U_{e4}|^2\sqrt{\Delta{m}^{2}_{41}}$
of $\nu_{4}$
to the effective Majorana mass in neutrinoless double-$\beta$ decay.
Figure~\ref{fig:mbb-chi}
shows the marginal $\Delta\chi^2$
as a function of $m_{\beta\beta}^{(4)}$
obtained from the fit of short-baseline oscillation data in Ref.~\cite{1210.5715}
(dash-dotted curve).
One can see that in this case $m_{\beta\beta}^{(4)}$ has no upper bound at 83\% CL.
The dashed curve in Fig.~\ref{fig:mbb-chi},
obtained from the combined fit of oscillation and Mainz data,
gives an upper bound for $m_{\beta\beta}^{(4)}$ of
0.91
at $2\sigma$.
The solid curve in Fig.~\ref{fig:mbb-chi},
obtained with the addition of Troitsk data,
improve dramatically the bound,
because the marginal
$\Delta\chi^2$
increases steeply for $m_{\beta\beta}^{(4)}$
larger than the best-fit value at about
$0.08 \, \text{eV}$.
Considering also the lower bound for $m_{\beta\beta}^{(4)}$ given by short-baseline oscillation data alone,
we obtain
\begin{equation}
0.013
\lesssim
m_{\beta\beta}^{(4)}
\lesssim
0.20
\, \text{eV}
\qquad
(2\sigma)
\,.
\label{mbb}
\end{equation}
The slight decrease of the marginal
$\Delta\chi^2$
for $m_{\beta\beta}^{(4)}$ smaller than the best-fit value
obtained with the inclusion of Troitsk data is due to a slight increase of the value of $\chi^2_{\text{min}}$,
from 45.5
to 45.8.

Considering the case in which the contribution
$m_{\beta\beta}^{(4)}$
to the effective Majorana mass is not canceled
by that of the three light neutrinos
(i.e. $m_{\beta\beta} \ge m_{\beta\beta}^{(4)}$;
see the discussion in Ref.~\cite{1210.5715} and Refs.~\cite{Barry:2011wb,1110.5795}),
Fig.~\ref{fig:mbb-chi} shows also
the currently most stringent 90\% CL upper bound for $m_{\beta\beta}$
obtained in Ref.~\cite{1211.3863}
from the combined
EXO \cite{Auger:2012ar}
and
KamLAND-Zen \cite{1211.3863}
data,
taking into account nuclear matrix element uncertainties.
One can see that this upper bound erodes the upper bound in Eq.~(\ref{mbb})
for large values of the nuclear matrix element.
The interesting range of $m_{\beta\beta}^{(4)}$
below the EXO+KamLAND-Zen upper bound will be explored
by several neutrinoless double-$\beta$ decay experiments in the near future
(see Refs.~\cite{1206.2560,1210.7432}).

Figure~\ref{fig:mbb-chi} shows also
the $1\sigma$
Klapdor-Kleingrothaus et al. range of $m_{\beta\beta}$
\cite{KlapdorKleingrothaus:2006ff}.
Besides being disfavored by the EXO+KamLAND-Zen upper bound on $m_{\beta\beta}$,
it is also disfavored by our results
if $m_{\beta\beta}^{(4)}$ is the dominant contribution
to the effective Majorana mass
and also if there is a cancellation of $m_{\beta\beta}^{(4)}$ with the contribution
to the effective Majorana mass of the three light neutrinos \cite{Barry:2011wb,1110.5795,1210.5715},
i.e. if $m_{\beta\beta} \le m_{\beta\beta}^{(4)}$.

In conclusion,
we have obtained an interesting upper bound for
$\Delta{m}^{2}_{41}$
in the framework of 3+1 mixing
from the results of short-baseline $\nu_{e}$ and $\bar\nu_{e}$ oscillation data
and from the recent results of a search for the effects of $m_{4}^2$
in the spectrum of the electrons emitted in Tritium decay
in the Troitsk nu-mass experiment.
The upper bound for
$\Delta{m}^{2}_{41}$
implies that it is likely that the electron neutrino oscillation length is sufficiently large
to measure the dependence of the disappearance probability
as a function of distance and/or energy
in electron neutrino and antineutrino radioactive source experiments.
It is also favorable for the proposed ICARUS@CERN experiment,
because it implies that it is likely that oscillations are not washed-out in the near detector.
We have also discussed the implications of our results for neutrinoless double-$\beta$ decay.

\bigskip
\centerline{\textbf{Acknowledgments}}
\medskip

We are very grateful to Vladislav Pantuev
for sending us the Troitsk likelihood data.
The work of Y.F. Li is supported in part by the National Natural Science Foundation
of China under grant No.~11135009.
The work of H.W. Long is supported in part by the National Natural Science Foundation
of China under grant No.~11265006.


\begin{thebibliography}{10}

\bibitem{Mueller:2011nm}
T.~A. Mueller {\em et~al.},
Phys. Rev. {\bf C83}, 054615 (2011), arXiv:1101.2663.

\bibitem{Huber:2011wv}
P.~Huber,
Phys. Rev. {\bf C84}, 024617 (2011), arXiv:1106.0687.

\bibitem{Mention:2011rk}
G.~Mention {\em et~al.},
Phys. Rev. {\bf D83}, 073006 (2011), arXiv:1101.2755.

\bibitem{Laveder:2007zz}
M.~Laveder,
Nucl. Phys. Proc. Suppl. {\bf 168}, 344 (2007),
{Workshop on Neutrino Oscillation Physics (NOW 2006), Otranto, Lecce,
Italy, 9-16 Sep 2006}.

\bibitem{hep-ph/0610352}
C.~Giunti and M.~Laveder,
Mod. Phys. Lett. {\bf A22}, 2499 (2007), hep-ph/0610352.

\bibitem{Giunti:2010zu}
C.~Giunti and M.~Laveder,
Phys. Rev. {\bf C83}, 065504 (2011), arXiv:1006.3244.

\bibitem{1210.5715}
C.~Giunti, M.~Laveder, Y.~Li, Q.~Liu, and H.~Long,
Phys. Rev. {\bf D86}, 113014 (2012), arXiv:1210.5715.

\bibitem{Giunti:2007ry}
C.~Giunti and C.~W. Kim,
{\em {Fundamentals of Neutrino Physics and Astrophysics}} (Oxford
University Press, Oxford, UK, 2007),
{ISBN 978-0-19-850871-7}.

\bibitem{Bilenky:2010zza}
S.~Bilenky,
{\em {Introduction to the physics of massive and mixed neutrinos}}
(Springer, 2010),
{Lecture Notes in Physics, Volume 817; ISBN 978-3-642-14042-6}.

\bibitem{Xing:2011zza}
Z.-z. Xing and S.~Zhou,
{\em {Neutrinos in particle physics, astronomy and cosmology}}
(Zhejiang University Press, 2011),
{ISBN 978-7-308-08024-8}.

\bibitem{hep-ex/0509008}
ALEPH, DELPHI, L3, OPAL, SLD, LEP Electroweak Working Group, SLD Electroweak
Group, SLD Heavy Flavour Group, S.~Schael {\em et~al.},
Phys. Rept. {\bf 427}, 257 (2006), hep-ex/0509008.

\bibitem{hep-ph/9812360}
S.~M. Bilenky, C.~Giunti, and W.~Grimus,
Prog. Part. Nucl. Phys. {\bf 43}, 1 (1999), hep-ph/9812360.

\bibitem{hep-ph/0405172}
M.~Maltoni, T.~Schwetz, M.~Tortola, and J.~Valle,
New J. Phys. {\bf 6}, 122 (2004), hep-ph/0405172.

\bibitem{hep-ph/0606054}
A.~Strumia and F.~Vissani,
(2006), hep-ph/0606054.

\bibitem{GonzalezGarcia:2007ib}
M.~C. Gonzalez-Garcia and M.~Maltoni,
Phys. Rept. {\bf 460}, 1 (2008), arXiv:0704.1800.

\bibitem{hep-ex/0104049}
LSND, A.~Aguilar {\em et~al.},
Phys. Rev. {\bf D64}, 112007 (2001), hep-ex/0104049.

\bibitem{1207.4809}
MiniBooNE, A.~A. Aguilar-Arevalo {\em et~al.},
(2012), arXiv:1207.4809.

\bibitem{1103.4570}
J.~Kopp, M.~Maltoni, and T.~Schwetz,
Phys. Rev. Lett. {\bf 107}, 091801 (2011), arXiv:1103.4570.

\bibitem{1107.1452}
C.~Giunti and M.~Laveder,
Phys.Rev. {\bf D84}, 073008 (2011), arXiv:1107.1452.

\bibitem{1109.4033}
C.~Giunti and M.~Laveder,
Phys.Rev. {\bf D84}, 093006 (2011), arXiv:1109.4033.

\bibitem{1111.1069}
C.~Giunti and M.~Laveder,
In Phys. Lett. \cite{Giunti:2011cp}, pp. 200--207, arXiv:1111.1069.

\bibitem{1207.4765}
J.~M. Conrad, C.~M. Ignarra, G.~Karagiorgi, M.~H. Shaevitz, and J.~Spitz,
(2012), arXiv:1207.4765.

\bibitem{Declais:1995su}
Bugey, B.~Achkar {\em et~al.},
Nucl. Phys. {\bf B434}, 503 (1995).

\bibitem{Declais:1994ma}
Bugey, Y.~Declais {\em et~al.},
Phys. Lett. {\bf B338}, 383 (1994).

\bibitem{Kuvshinnikov:1990ry}
A.~Kuvshinnikov, L.~Mikaelyan, S.~Nikolaev, M.~Skorokhvatov, and A.~Etenko,
JETP Lett. {\bf 54}, 253 (1991).

\bibitem{Zacek:1986cu}
CalTech-SIN-TUM, G.~Zacek {\em et~al.},
Phys. Rev. {\bf D34}, 2621 (1986).

\bibitem{Hoummada:1995zz}
A.~Hoummada, S.~Lazrak~Mikou, G.~Bagieu, J.~Cavaignac, and D.~Holm~Koang,
Applied Radiation and Isotopes {\bf 46}, 449 (1995).

\bibitem{Vidyakin:1990iz}
Krasnoyarsk, G.~S. Vidyakin {\em et~al.},
Sov. Phys. JETP {\bf 71}, 424 (1990).

\bibitem{Anselmann:1995ar}
GALLEX, P.~Anselmann {\em et~al.},
Phys. Lett. {\bf B342}, 440 (1995).

\bibitem{Hampel:1998fc}
GALLEX, W.~Hampel {\em et~al.},
Phys. Lett. {\bf B420}, 114 (1998).

\bibitem{1001.2731}
F.~Kaether, W.~Hampel, G.~Heusser, J.~Kiko, and T.~Kirsten,
Phys. Lett. {\bf B685}, 47 (2010), arXiv:1001.2731.

\bibitem{Abdurashitov:1996dp}
SAGE, J.~N. Abdurashitov {\em et~al.},
Phys. Rev. Lett. {\bf 77}, 4708 (1996).

\bibitem{hep-ph/9803418}
SAGE, J.~N. Abdurashitov {\em et~al.},
Phys. Rev. {\bf C59}, 2246 (1999), hep-ph/9803418.

\bibitem{nucl-ex/0512041}
J.~N. Abdurashitov {\em et~al.},
Phys. Rev. {\bf C73}, 045805 (2006), nucl-ex/0512041.

\bibitem{0901.2200}
SAGE, J.~N. Abdurashitov {\em et~al.},
Phys. Rev. {\bf C80}, 015807 (2009), arXiv:0901.2200.

\bibitem{Frekers:2011zz}
D.~Frekers {\em et~al.},
Phys. Lett. {\bf B706}, 134 (2011).

\bibitem{Giunti:2009xz}
C.~Giunti and Y.~Li,
Phys. Rev. {\bf D80}, 113007 (2009), arXiv:0910.5856.

\bibitem{Palazzo:2011rj}
A.~Palazzo,
Phys. Rev. {\bf D83}, 113013 (2011), arXiv:1105.1705.

\bibitem{Palazzo:2012yf}
A.~Palazzo,
Phys. Rev. {\bf D85}, 077301 (2012), arXiv:1201.4280.

\bibitem{Palazzo-NOW2012}
A.~Palazzo,
(2012),
{NOW 2012, Neutrino Oscillation Workshop, 9-16 September 2012, Conca
Specchiulla, Otranto, Italy}.

\bibitem{Bodmann:1994py}
KARMEN., B.~E. Bodmann {\em et~al.},
Phys. Lett. {\bf B332}, 251 (1994).

\bibitem{hep-ex/9801007}
KARMEN, B.~Armbruster {\em et~al.},
Phys. Rev. {\bf C57}, 3414 (1998), hep-ex/9801007.

\bibitem{hep-ex/0105068}
LSND, L.~B. Auerbach {\em et~al.},
Phys. Rev. {\bf C64}, 065501 (2001), hep-ex/0105068.

\bibitem{1106.5552}
J.~Conrad and M.~Shaevitz,
Phys. Rev. {\bf D85}, 013017 (2012), arXiv:1106.5552.

\bibitem{Giunti:2011cp}
C.~Giunti and M.~Laveder,
Phys. Lett. {\bf B706}, 200 (2011), arXiv:1111.1069.

\bibitem{Wong:2011ip}
Y.~Y.~Y. Wong,
Ann. Rev. Nucl. Part. Sci. {\bf 61}, 69 (2011), arXiv:1111.1436.

\bibitem{Abazajian:2012ys}
K.~N. Abazajian {\em et~al.},
(2012), arXiv:1204.5379.

\bibitem{Steigman:2012ve}
G.~Steigman,
Adv. High Energy Phys. {\bf 2012}, 268321 (2012), arXiv:1208.0032.

\bibitem{1210.4194}
C.~Kraus, A.~Singer, K.~Valerius, and C.~Weinheimer,
(2012), arXiv:1210.4194.

\bibitem{1211.7193}
A.~I. Belesev {\em et~al.},
(2012), arXiv:1211.7193.

\bibitem{1211.3863}
KamLAND-Zen, A.~Gando {\em et~al.},
Phys. Rev. Lett. {\bf 110}, 062502 (2013), arXiv:1211.3863.

\bibitem{KlapdorKleingrothaus:2006ff}
H.~V. Klapdor-Kleingrothaus and I.~V. Krivosheina,
Mod. Phys. Lett. {\bf A21}, 1547 (2006).

\bibitem{hep-ex/9901012}
A.~Ianni, D.~Montanino, and G.~Scioscia,
Eur. Phys. J. {\bf C8}, 609 (1999), hep-ex/9901012.

\bibitem{1006.2103}
V.~N. Gavrin, V.~V. Gorbachev, E.~P. Veretenkin, and B.~T. Cleveland,
(2010), arXiv:1006.2103.

\bibitem{1011.4509}
S.~K. Agarwalla and R.~S. Raghavan,
(2010), arXiv:1011.4509.

\bibitem{1107.2335}
M.~Cribier {\em et~al.},
Phys. Rev. Lett. {\bf 107}, 201801 (2011), arXiv:1107.2335.

\bibitem{1109.6036}
D.~Dwyer, K.~Heeger, B.~Littlejohn, and P.~Vogel,
(2011), arXiv:1109.6036.

\bibitem{1110.2983}
Y.~Novikov {\em et~al.},
(2011), arXiv:1110.2983.

\bibitem{Ianni-NOW2012}
A.~Ianni,
(2012),
{NOW 2012, Neutrino Oscillation Workshop, 9-16 September 2012, Conca
Specchiulla, Otranto, Italy}.

\bibitem{Link-NOW2012}
J.~Link,
(2012),
{NOW 2012, Neutrino Oscillation Workshop, 9-16 September 2012, Conca
Specchiulla, Otranto, Italy}.

\bibitem{Antonello:2012hf}
M.~Antonello {\em et~al.},
(2012), arXiv:1203.3432.

\bibitem{Stanco-NOW2012}
L.~Stanco,
(2012),
{NOW 2012, Neutrino Oscillation Workshop, 9-16 September 2012, Conca
Specchiulla, Otranto, Italy}.

\bibitem{Gibin-NOW2012}
D.~Gibin,
(2012),
{NOW 2012, Neutrino Oscillation Workshop, 9-16 September 2012, Conca
Specchiulla, Otranto, Italy}.

\bibitem{1204.2449}
A.~V. Derbin, A.~S. Kayunov, and V.~N. Muratova,
(2012), arXiv:1204.2449.

\bibitem{1205.2955}
A.~P. Serebrov {\em et~al.},
(2012), arXiv:1205.2955.

\bibitem{Gaffiot-NOW2012}
J.~Gaffiot,
(2012),
{NOW 2012, Neutrino Oscillation Workshop, 9-16 September 2012, Conca
Specchiulla, Otranto, Italy}.

\bibitem{1212.2182}
K.~M. Heeger, B.~R. Littlejohn, H.~P. Mumm, and M.~N. Tobin,
(2012), arXiv:1212.2182.

\bibitem{1205.4419}
A.~Bungau {\em et~al.},
(2012), arXiv:1205.4419.

\bibitem{Riis:2010zm}
A.~S. Riis and S.~Hannestad,
JCAP {\bf 1102}, 011 (2011), arXiv:1008.1495.

\bibitem{SejersenRiis:2011sj}
A.~S. Riis, S.~Hannestad, and C.~Weinheimer,
Phys. Rev. {\bf C84}, 045503 (2011), arXiv:1105.6005.

\bibitem{Formaggio:2011jg}
J.~A. Formaggio and J.~Barrett,
Phys. Lett. {\bf B706}, 68 (2011), arXiv:1105.1326.

\bibitem{Esmaili:2012vg}
A.~Esmaili and O.~L.~G. Peres,
Phys. Rev. {\bf D85}, 117301 (2012), arXiv:1203.2632.

\bibitem{Barry:2011wb}
J.~Barry, W.~Rodejohann, and H.~Zhang,
JHEP {\bf 07}, 091 (2011), arXiv:1105.3911.

\bibitem{1110.5795}
Y.~Li and S.~Liu,
Phys. Lett. {\bf B706}, 406 (2012), arXiv:1110.5795.

\bibitem{Auger:2012ar}
EXO Collaboration, M.~Auger {\em et~al.},
(2012), arXiv:1205.5608.

\bibitem{1206.2560}
W.~Rodejohann,
J. Phys. {\bf G39}, 124008 (2012), arXiv:1206.2560.

\bibitem{1210.7432}
B.~Schwingenheuer,
(2012), arXiv:1210.7432.

\end{thebibliography}

\end{document}